# High-fidelity velocity and concentration measurements of turbulent buoyant jets

Valentina Valori,[a,b]* Sunming Qin[c], Victor Petrov[a,b,d] and Annalisa Manera[a,b,d]

[a]*ETH Zürich, Dep. of Mechanical and Process Eng., Nuclear Safety & Multiphase Flows, Sonneggstrasse 3, 8092 Zürich, Switzerland*

[b]*Paul Scherrer Institute, Laboratory for Reactor Physics and Thermal-Hydraulics, Forschungsstrasse 111, 5232 Villigen PSI, Switzerland*

[c]*Idaho National Laboratory, 1955 N. Fremont Ave. Idaho Falls, ID 83415*

[d]*University of Michigan, Department of Nuclear Engineering & Radiological Sciences, 2355 Bonisteel Boulevard, Ann Arbor, Michigan 48105*

*E-mail: vvalori@ethz.ch

# High-fidelity velocity and concentration measurements of turbulent buoyant jets

Accurate models of turbulent buoyant flows are essential for the design of nuclear reactors thermal hydraulics and passive safety systems. However, available models fail to fully capture the physics of turbulent mixing when buoyancy becomes predominant with respect to momentum. Therefore, high-fidelity experiments of well-controlled fundamental flows are needed to develop and validate more accurate models. We analyze experiments of positive and negative turbulent buoyant jets, both in uniform and stratified environments, with the aim of understanding the thermal hydraulics of turbulent mixing with variable density and providing high-fidelity data for the development and validation of turbulence models. Non-intrusive, simultaneous Particle Image Velocimetry and Laser Induced Fluorescence measurements were carried out to acquire instantaneous velocity and concentration fields on a vertical section parallel to the axis of a jet in the self-similar region. The Refractive Index Matching method was applied to measure high-resolution buoyant jets with up to 8.6% density difference. These data are free of the typical errors that characterize optical measurements of buoyancy driven flows (e.g., natural and mixed convection) where the refractive index of the fluid is inhomogeneous throughout the measurement domain. Turbulent statistics and entrainment of buoyant jets in uniform and stratified environments are presented. These data are compared with non-buoyant jets in uniform environment, as a reference to investigate the effects of buoyancy and stratification on turbulent mixing. The results will be used for the assessment of current turbulence models and as basis for the development of a new one that captures turbulent mixing.

## I. INTRODUCTION

Turbulent buoyant jets in density stratified environments are ubiquitous in geophysical and technological applications. Some examples are the dispersion of pollutants in the atmosphere or oil spills in the ocean. Density effects like fluid stratification play a significant role in nuclear power plants [1], especially in the design of safety systems. In case of severe accidents in light water reactors, hydrogen may be generated leading to the formation of a layer of flammable and explosive gas that may ignite if a critical concentration with oxygen is locally reached. Due to the effect of buoyancy, hydrogen will accumulate as a stratified layer at the top of the containment. In the Three Mile Island accident (1979) and in the Fukushima Daiichi (2011) one, a loss of core cooling caused a reaction between water and zircaloy that generated a safety-relevant amount of hydrogen. Several safety features are currently applied to avoid thermal stratification of hydrogen, like sprays injections, catalytic recombination and gas jets. Also, passive systems that lead to buoyancy driven flows may promote mixing without external momentum sources [2]. Thermal stratification is a safety issue also in Generation IV Lead Cooled Fast Reactors, where fluid stratification in the pools may cause safety concerns regarding the structural integrity of the reactor, due to thermal fatigue of the components [3]. The present work focuses on the study of turbulent buoyant jets in uniform and stratified environments based on advanced optical measurements.

When a fluid is injected in an environment with different density, both the pressure drops through the orifice and buoyancy contribute to its motion. If the jet is lighter than the environment, the two forces act in the same direction (positive buoyant jet). If instead the jet is heavier than the environment, buoyancy acts in the opposite direction (negative buoyant jet). When a fluid is injected in a two-layer stratified environment with heavier fluid at the bottom (see for example Figure 2, right panel), the momentum force is enhanced by the effect of buoyancy in the bottom layer, while it is contrasted in the top one. If the momentum is not large enough, the jet may behave like a fountain and reverses its direction at the interface. Buoyant jets in uniform and stratified environment have been investigated in the past both theoretically and through computations and experiments [4, 5, 6, 7, 8, 9, 10, 11, 2, 12]. Ricou and Spalding [11] studied axisymmetric buoyant jets in uniform environment by measuring the jet's entrainment through the axial mass flow rate. They modelled the entrainment rate dependence on the jet density, predicting that the entrainment coefficient is reduced as the jet's density becomes smaller with respect to the one of the environment. Turner [5] studied the behaviour of negative buoyant jets (also called fountains) with qualitative flow visualizations and determined their penetration depth. More recently, Viggiano et al. [13] studied turbulence instabilities of buoyant jets considering the entrainment and turbulent fluctuations obtained from Particle Image Velocimetry (PIV) measurements. A combined approach of PIV experiments, theory and Large-Eddy Simulations (LES) was used by Salizzoni et al. [14] to study the entrainment of turbulent buoyant jets. They concluded that the density ratio between the jet and the environment modifies the turbulent kinetic energy production and the shape of the mean velocity profile, but does not affect the entrainment, which does not confirm previous studies (e.g. [11]). Further research is necessary to determine the dependence of the entrainment on the density ratio between the jet and the environment. Flow visualizations using a dye in stratified environment were made by Lin and Linden [10] to determine penetrative entrainment rates of turbulent fountains, by Ansong et al. [15] to study the penetration depth of a water fountain, and by Shakouchi [16] to investigate the jet impingement on a stratification interface depending on the Reynolds number. Mott and Woods [17] studied plumes theoretically and experimentally applying flow visualizations and conductivity probes. Deri et. al. [2] studied the mixing of air fountains in stratified environment with conductivity sensors and PIV. Herault et al. [18] performed simultaneous velocity and density measurements, with PIV and LIF respectively, to study the impinging of a jet on a density interface. They observed that the turbulent mass flux is almost perpendicular to the density gradient at the stratification interface, which is in contrast with the assumptions made in the eddy diffusivity model used in numerical schemes like in the Reynolds Averaged Navier-Stokes (RANS) equations. Computational Fluid Dynamics (CFD) codes currently cannot resolve all the scales of complex flows typical of engineering applications, because of limitations in the computational power. Turbulence models are therefore employed to determine the smallest scales of the flow. A spatial filtering is applied in the Large Eddy Simulation model, while a temporal filtering is used in the RANS model. In LES, subgrid-scale models that assume isotropy at small scales are applied. This assumption is not valid for turbulent buoyant jets in stratified environment, and therefore LES simulations, besides having high computational requirements, fail to correctly reproduce the physics of these flows [19]. In RANS models, the time dependent quantities of the Navier-Stokes equations are represented by a time-averaged and a fluctuating component. For example, the instantaneous velocity field is represented as: $\boldsymbol{u}_j = \bar{u}_j + u'_j$. Using Einstein notation, the NS equations can be written as:

$$\frac{\partial \rho}{\partial t} + \frac{\partial (\rho \bar{u}_j)}{\partial x_j} = 0, \qquad 1)$$

$$\frac{\partial (\rho \bar{u}_i)}{\partial t} + \frac{\partial}{\partial x_j}(\rho \overline{u_i u_j}) = \frac{\partial \rho}{\partial x_i} + \frac{\partial}{\partial x_j}\left[\eta\left(\frac{\partial \bar{u}_i}{\partial x_j} + \frac{\partial \bar{u}_j}{\partial x_i}\right) - \rho\overline{u'_i u'_j}\right] + g_j(\rho - \rho_0), \qquad 2)$$

$$\frac{\partial (\rho \bar{c})}{\partial t} + \frac{\partial}{\partial x_j}\left(\Gamma \frac{\partial c}{\partial x_j}\right) - \frac{\partial}{\partial x_j}(\rho\overline{u'_j c'}), \qquad 3)$$

where $\rho$ is the density field, $c$ the concentration scalar field, $\eta$ the dynamic viscosity, and $\Gamma$ the diffusivity.

To solve this set of equations, one needs to model the cross correlations of the velocity fluctuations $\overline{u'_i u'_j}$ (known as the Reynolds shear stresses) as well as the turbulent fluxes $\overline{u'_j c'}$. In this way the number of unknowns will be equal to the number of equations. Petrov and Manera [19] pointed out that current models for turbulent momentum fluxes have the weak point of overestimating the thermal stratification term and underestimating turbulent mixing, when buoyancy prevails over momentum. To improve our knowledge of the physics of turbulent buoyant jets in stratified environments and validate current turbulence models, experimental data at high spatial resolution are required.

PIV and LIF allow us to obtain planar velocity and concentration fields on a whole cross-section of the domain at high spatial resolution in a non-intrusive way. However, the accuracy of these optical techniques may be challenged by refractive index inhomogeneities when dealing with buoyant flows [20, 21, 22]. Density differences are indeed associated with refractive index differences that may cause errors in the PIV measurements for example regarding blurring of the particles, the determination of their position and velocity, and the location and shape of the light sheet [23]. This may impede measurement with high-spatial resolution. It is possible to overcome the problem of optical distortions when studying mixing with variable density, by accurately matching the refractive index of the fluids used for the experiments. Krohn et al. [24] proposed a Refractive Index Matching (RIM) method that allowed to reach a density difference of up to 8.6% while keeping the refractive index of the two solutions constant even during the mixing. The application of this method allows us to study the mixing of solutions with different densities without having optical distortions in the fluid. Qin et al. [25] applied the RIM method to study the mixing of turbulent buoyant jets in uniform environment with a density difference of 3.16%. They acquired planar velocity and concentration fields at high spatial and temporal resolution. They observed an increase of mixing for positive buoyant jets with respect than for neutrally buoyant jets at the same Reynolds number, and a suppression of mixing for negative buoyant jets with respect to the reference case with neutral buoyancy. In reference [26] Qin et al. applied the RIM method to measure the interaction of turbulent buoyant jets with two layers stratified environment with a density difference of 3.16%.

The goal of the present study is to investigate the effect of buoyancy and of a two-layer stratified environment on the entrainment and turbulent statistics of turbulent buoyant jets. The analysis is based on high-resolution PIV and planar LIF measurements with a density difference of up to 8.6% where the RIM method was applied. The effect of buoyancy is studied comparing cases of positive, neutral and negative buoyant jets. The cases with a two-layers stratified environment are compared with cases in uniform environment (see Figure 2 for a representation of uniform and stratified environment). The experiments presented are summarized in Table 1. Cases with different level of turbulence, expressed by the Reynolds number, and of buoyancy with respect to shear, expressed by the Richardson number are analyzed.

The Reynolds number along the jet's axis is defined as:

$$Re(x_0, y) = \frac{DV(x_0, y)}{v(x_0, y)}, \qquad 4)$$

where $D$ is the diameter of the jet's orifice, $V(x_0, y)$ is the time-averaged vertical velocity on the axis of the jet $(x = x_0, y)$. and $v$ is the kinematic viscosity of the fluid.

The Richardson number is expressed as:

$$Ri = gD\frac{\rho_{env.} - \rho_{jet}}{\rho_{jet} V_0^2}, \qquad 5)$$

where $g$ is the magnitude of the vertical acceleration, $\rho_{env.}$ is the mass density of the environment where the jet is injected, $\rho_{jet}$ is the mass density of the jet at the orifice, and $V_0$ is the jet's velocity at the orifice. Mean quantities like the local Reynolds number and entrainment, and turbulent statistics like Reynolds stresses and concentration fluxes are shown.

**I.A. Methods**

The jet's development, turbulent Reynolds stresses, concentration fluxes and entrainment, are experimentally studied in (buoyant) jets with uniform and stratified environments. The effect of buoyancy is studied by mixing fluids with different densities and the same refractive index. The novel Refractive Index Matching (RIM) method [24], which was developed at the Experimental and Computational Multiphase Flow Laboratory of the University of Michigan, allows us to measure with high accuracy the mixing of fluids up to a density difference of 8.6% extending the capabilities of available RIM methods [27, 28, 29]. The novel approach is based on the behavior of excess properties and their variability while mixing two ternary solutions. With this technique, one can study buoyancy without heating the fluid. When a large density difference (e.g., above 5-10%) is obtained by heating a fluid, the results of the measurements are affected by non-Oberbeck-Boussinesq effects [30, 31], because the approximation of constant fluid properties and density that depends linearly on temperature only in the buoyancy term is not applicable. The mixing of fluids with non-uniform refractive index causes deviations of the laser light and of the light scattered by the seeding particles [21, 23, 20]. Both effects challenge PIV and LIF experiments. With the RIM method, one can use optical techniques with high accuracy. In the present work simultaneous LIF and PIV measurements are used to study the mixing of buoyant jets with a stagnant environment. The environment may be uniform or with a two-layer stratification, For LIF measurements, the jet's fluid is marked with a tracer compound, while the environment contains clean fluid. In case of two layers stratification, the upper layer of the environment contains traced fluid.

In the experiments with 3.16% density difference, a combination of two aqueous solutions with sodium sulphate ($Na_2SO_4$) and glycerol ($H_2O - CH_2OHCHOHCH_2OH$) was used. In the measurements with 8.6% density difference, the liquids used were ternary solutions of water, isopropanol ($CH_3CHOHCH_3$), and glycerol. All three liquids are fully miscible among each other, and the two solutions exhibit a large density difference. Velocity and concentration measurements are performed in the self-similar region of the jet. The raw data used in this work were published in the Ph.D. thesis of S. Qin [32].

## II. EXPERIMENTAL SET-UP

The experimental set-up and instrumentation are described in detail in reference [32], we report here a summary of the most important characteristics. The miniDESTROJER (DEnsity Stratified Turbulent ROund free Jet ExpeRiment) facility at the University of Michigan is a facility designed to study turbulent buoyant jets in stratified environments using optical measurements (a vertical sketch of the facility is shown in Figure 1). It is composed of a tank with glass walls with a nozzle of 2 mm diameter at the bottom. The tank size is 300 mm × 300 mm × 300 mm. The corresponding tank size to nozzle diameter ratio is 150, ensuring the flow is an unconfined free jet. The axial jet flow is pumped with a piston powered by a Dynamic Motor Motion Servo engine that guarantees an accurate control of the inlet jet's velocity. A scheme of the miniDESTROJER test section for experiments with uniform and stratified environment is shown in Figure 2.

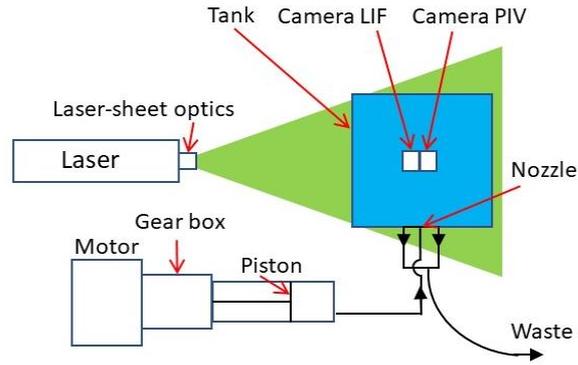

Figure 1: Vertical sketch of the experimental set-up composed of the MiniDESTROJER facility and the instrumentation for PIV amd LIF measurements.

Planar PIV is used to measure instantaneous velocity fields on a vertical section of the jet in the self-similar region, at a downstream location from 40 to 80 orifice diameters. The RIM method was applied to measure buoyant jets with density differences up to 8.6%. The instrumentation for the PIV measurements consists of one high-speed laser with optics, one high-speed camera, a synchronizer, and a software for data acquisition and processing. The highspeed double pulsed laser is a Nd:YLF laser from Photonics Industries International Inc. with a maximum pulse frequency of 10 kHz at a wavelength of 527 nm and 70 mJ energy per pulse. It is equipped with two spherical and one cylindrical lens, with focal length $f = 10$mm, to create a laser sheet. The camera is a Phantom Miro LAB M340 highspeed CMOS with $2560 \times 1600$ pixels, pixel size of 10 µm, and 800 Hz maximum frame rate at full resolution. It was equipped with a Nikon Nikkor AF lens with focal length 50 mm. The seeding particles are hollow glass spheres with diameter of 10 μm and mass density of 1.10±0.05 g/cm$^3$. The software DaVis 8.4 from LaVision® GmbH was used for data acquisition and processing. A spatial resolution of $818 \times 818$ µm$^2$ was reached [32].

For the LIF measurements, Rhodamine-6G was used as tracer to characterize the scalar field of the jet flow. The laser light at a wavelength of 532 nm, acts on the tracer molecules exciting and causing the same to emit fluorescent signals. Images are captured by a CMOS digital camera equipped with a light filter permeable to the particle's emission wavelength of 570 nm. The PIV and LIF measurements are synchronized to have simultaneous velocity and concentration fields on the same vertical section of the jet.

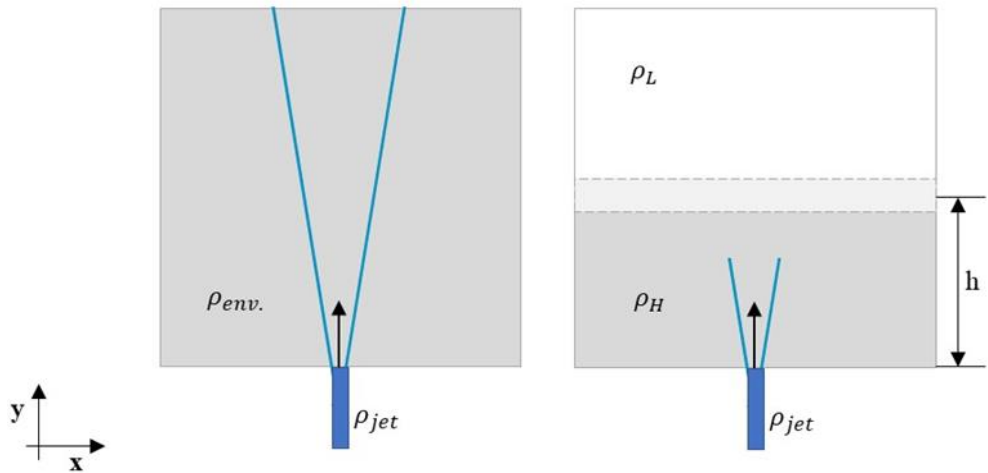

Figure 2: Scheme of the experiments of jets with uniform (left) or stratified environment (right). $\rho_{jet}$ and $\rho_{env.}$ are the densities of the jet and the environment fluid, respectively. $\rho_H$ and $\rho_L$ are the densities of the heavy and the light fluid, respectively. $h$ is the vertical distance from the jet orifice. $x$ and $y$ are the horizontal and vertical axis, respectively.

Turbulent buoyant jets in density stratified environments are ubiquitous in geophysical and technological applications. Some examples are the dispersion of pollutants in the atmosphere or oil spills in the ocean.

## II. EXPERIMENTAL PROGRAMME

The experimental test matrix includes cases of turbulent buoyant jets in uniform and stratified environment, and reference cases of jets in uniform environment. Table 1 reports the experimental conditions of all cases studied, the jet's velocity at the orifice ($V_0$), the mass density of the jet and the one of the environment ($\rho_{jet}$ and $\rho_{env}$, respectively), the temperature of the set-up (T), the Reynolds and Richardson numbers of the jet, defined as in equations (4) and (5), respectively. The cases investigated can be grouped in four categories:

- Positive buoyant jets in uniform environment, when the density of the jet is lighter than the density of the environment ($\rho_{jet} < \rho_{env}$). Cases D033 and D034 are characterized by a relative density difference between the environment and the jet of 3.16 % $\left(\frac{\rho_{jet} - \rho_{env.}}{\rho_{jet}} = 3.16\%\right)$, while cases D051 and D052 by a relative density difference of 8.6%.

- Negative buoyant jets in uniform environment, when the density of the jet is heavier than the density of the environment ($\rho_{jet} > \rho_{env}$). Cases D035 and D036 are negative buoyant jets with a negative density difference of 3.16%, while cases D047 and D048 have a negative density difference of 8.6%.

- Positive buoyant jets in stratified environment. Cases D042 and D043 have a density difference with respect to the environment (bottom layer) of 3.16%. As shown in Figure 2, right side, the environment is stratified, at the bottom there is heavy fluid ($\rho_{jet} < \rho_H$), and at the top (above the height of $h$) there is the same fluid of the jet ($\rho_{jet} = \rho_L$).

- Jets in uniform environment. These cases are used as a reference of no buoyancy. They are case D029 ÷ D032, D045, D046, D049, and D050.

Table 1: Summary of experiments in uniform environment without buoyancy (D029 ÷ D032, D045, D046, D049, D050), in uniform and stratified environment with relative density difference (Δρ) of 3.16% (D033 ÷ D036, D042 and D043), and in uniform environment with Δρ of 8.6% (D047, D048, D051, and D052). $V_0$ is the jet's velocity at the orifice. $\rho_{jet}$ and $\rho_{env}$ are the mass density of the jet and the environment, respectively. T is the temperature of the set-up. Re and Ri are the Reynolds and Richardson's numbers of the flow, defined in eq. (4) and (5), respectively.

| Case | [m/s] | $\rho_{jet}$ [kg/m³] | $\rho_{env}$ [kg/m³] | Δρ [%] | T [°C] | Re | Ri [×10⁻⁵] |
|---|---|---|---|---|---|---|---|
| **D029** | 5.990 ± 0.004 | 1012.0 ± 10.1 | | 0 | 20.4 ± 0.1 | 10000 | |
| **D030** | 2.460 ± 0.002 | 1012.0 ± 10.1 | | 0 | 20.4 ± 0.1 | 4000 | |
| **D031** | 5.530 ± 0.004 | 1044.0 ± 10.4 | | 0 | 20.0 ± 0.1 | 10000 | |
| **D032** | 2.300 ± 0.002 | 1044.0 ± 10.4 | | 0 | 20.0 ± 0.1 | 4000 | |
| **D045** | 9.060 ± 0.006 | 1011.0 ± 10.1 | | 0 | 19.2 ± 0.1 | 4000 | |
| **D046** | 4.220 ± 0.003 | 1011.0 ± 10.1 | | 0 | 19.3 ± 0.1 | 2000 | |
| **D049** | 9.060 ± 0.006 | 924.3 ± 9.2 | | 0 | 18.4 ± 0.1 | 4000 | |
| **D050** | 4.220 ± 0.003 | 924.3 ± 9.2 | | 0 | 18.6 ± 0.1 | 2000 | |
| **D033** | 5.990 ± 0.004 | 1012.0 ± 10.1 | 1044.0 ± 10.4 | 3.16 | 20.4 ± 0.1 | 10000 | 1.73 |
| **D034** | 2.460 ± 0.002 | 1012.0 ± 10.1 | 1044.0 ± 10.4 | 3.16 | 20.4 ± 0.1 | 4000 | 10.3 |
| **D035** | 5.530 ± 0.004 | 1044.0 ± 10.4 | 1012.0 ± 10.1 | 3.16 | 20.0 ± 0.1 | 10000 | -1.97 |
| **D036** | 2.300 ± 0.002 | 1044.0 ± 10.4 | 1012.0 ± 10.1 | 3.16 | 20.5 ± 0.1 | 4000 | -11.4 |
| **D042** | 5.990 ± 0.004 | 1012.0 ± 10.1 | Stratification | | 18.1 ± 0.1 | 10000 | |
| **D043** | 2.460 ± 0.002 | 1012.0 ± 10.1 | Stratification | | 19.0 ± 0.1 | 4000 | |
| **D047** | 9.060 ± 0.006 | 1011.0 ± 10.1 | 924.3 ± 9.2 | 8.60 | 19.3 ± 0.1 | 4000 | -2.05 |
| **D048** | 4.220 ± 0.003 | 1011.0 ± 10.1 | 924.3 ± 9.2 | 8.60 | 19.7 ± 0.1 | 2000 | -9.45 |
| **D051** | 9.060 ± 0.006 | 924.3 ± 9.2 | 1011.0 ± 10.1 | 8.60 | 18.5 ± 0.1 | 4000 | +2.00 |
| **D052** | 4.220 ± 0.003 | 924.3 ± 9.2 | 1011.0 ± 10.1 | 8.60 | 18.8 ± 0.1 | 2000 | +10.3 |

## III. RESULTS

Vertical mean velocity profiles at the axis location ($y/D = 40$) normalized by the initial velocity ($V_0$) for all the experiments considered (see Table 1) are shown in Figure 3 with errorbars according to the uncertainty estimation of Qin [32]. We can observe a collapse of the radial velocity profile for all the experiments without influence of Reynolds and Richardson numbers. This confirms the results of Hrycak et al. [33], which showed that for an axisymmetric turbulent jet, the radial velocity profile when normalized by the centerline velocity, is mostly independent of the Reynolds number. The effect of the Richardson number is not evident from this picture.

The Full Width at Half Height (FWHH) of the vertical velocity profile has the same slope for all experiments, except for case D043 (Figure 4). This experiment, together with case D042, is characterized by a two-layer stratified environment with a sharp interface (see Table 1). It has a lower Reynolds number than case D042 (Re = 4000, instead than Re = 10000) and therefore less momentum force. This is the reason why the jet's width after the interface, at approximately $\frac{y}{D} = 60$, starts to deviate from the one of the other experiments. The mean velocity on the jet's axis normalized by the velocity at the orifice ($V_m/V_0$) has a similar slope for all experiments, except for case D043 (Figure 5). The speed of this jet starts to decrease drastically at about $\frac{y}{D} = 70$. The slope of $V_m/V_0$ as a function of distance from the jet's orifice along the axis is, for the cases without stratification and for the one with stratification and large Reynolds number, in good agreement with the reference value reported in the literature from previous experiments in turbulent jets [7]. The results of Figure 4 and Figure 5 are explained by the fact that in the experiments with a stratified environment, when Re = 10000, as in case D042, the turbulent jet has enough momentum to penetrate through the top fluid surface. However, when the Reynolds number decreases as in case D043, the jet interacting with the density interface behaves like a fountain: it reaches a final maximum height, and then falls back to the density interface, interplaying with the layer and the upcoming flow.

The inverse of the mean tracer concentration on the jet's axis multiplied by the concentration at the orifice ($C_0/C_m$) shows a similar slope for all jets with or without buoyancy (Figure 6),which agrees quite well with the one from past studies from the literature [7]. The concentration profile along the axis was not plotted for the experiments in stratified environment, because in this case the upper layer of the environment was seeded to visualize the interface.

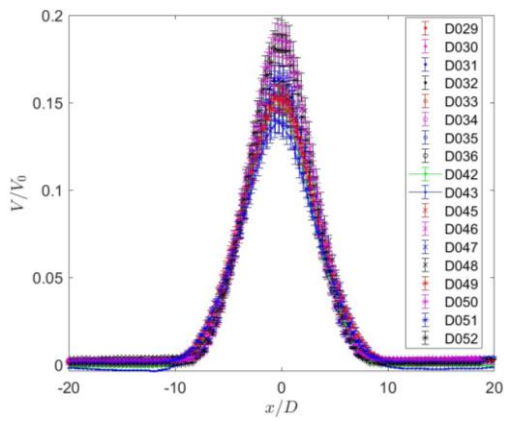 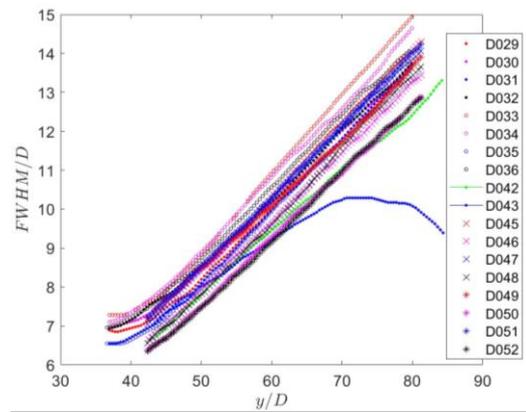

Figure 3: Spanwise velocity at $\frac{y}{D} = 40$, with errorbars.

Figure 4: Full Width at Half Maximum (FWHM) of the mean velocity profile.

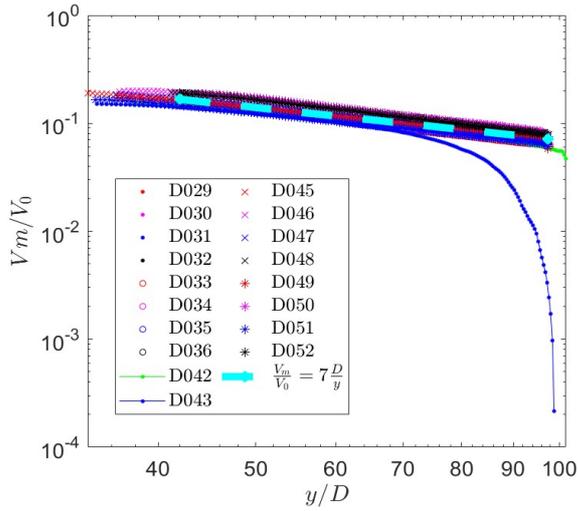
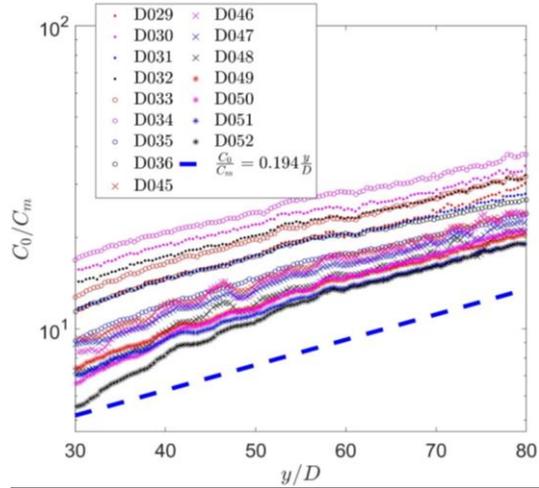

Figure 5: Mean velocity on the axis of a turbulent jet as a function of distance from the jet orifice.

Figure 6: Mean concentration on the axis of a turbulent jet as a function of distance from the orifice.

The evolution of the Reynolds number along the jet's axis for positive and negative buoyant jets with different speeds at the orifice is shown in Figure 7. The local Reynolds number and the Richardson number at the nozzle are computed as in equations (4) and (5), respectively. The kinematic viscosity used in the local Reynolds number is a function of the tracer concentration measured from the LIF experiments [25]. From Figure 7 we can observe that the Reynolds number along the jet's axis depends on the value of the Reynolds and Richardson numbers at the orifice. Indeed, cases D033 and D035 that are characterized by a larger initial Reynolds number ($Re = 10000$), have larger Reynolds along the axis with respect to cases D034, D036, and D051 ($Re = 4000$), and D052 ($Re = 2000$). Cases D033 and D034 that have a positive initial Richardson number ($Ri_0$) show larger values of Reynolds along the axis with respect to the corresponding cases at equal initial Reynolds number (D035, and D036 and D051, respectively). This confirms what was observed by Nathan [34] that only the jet inlet conditions will influence both the near-field and downstream flows, because they contribute to the structures of the turbulent motions that are carried from the jet nozzle throughout the flow. Cases D047 and D048, negative buoyant jets at low Reynolds number, were not included in the plot, because they may be affected by instability due to the balance of negative buoyancy and upward momentum flux [35, 36, 37, 38]. Figure 8 shows the data of Figure 7 normalized by the Reynolds number at the orifice ($Re_0$). We can observe that after normalization the values of Reynolds along the axis are ordered according to $Ri$, which means that buoyancy has a positive effect on the jet's speed along the axis. Figure 9 shows in addition errorbars for the same data.

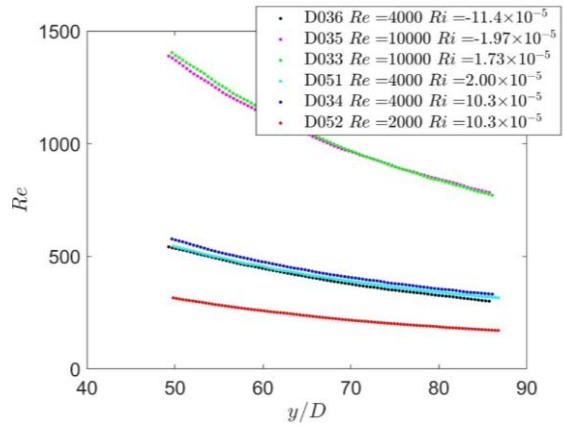

Figure 7: Reynolds number at the jet's centerline for experiments D033, D034, D035, D036, D051 and D052.

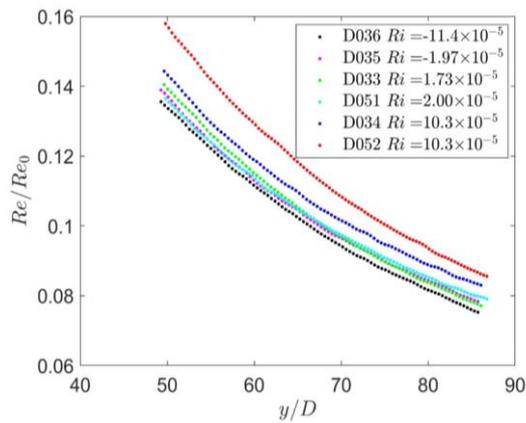 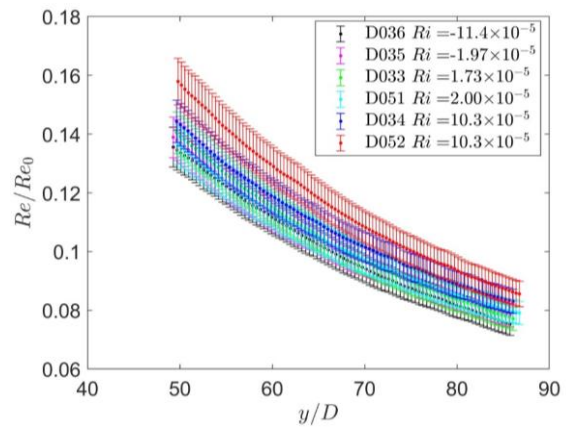

Figure 8: Reynolds number at the jet's    Figure 9: Reynolds number at the jet's

centerline normalized by the Richardson number at the orifice, for experiments D033, D034, D035, D036, D051 and D052.

centerline with errorbars normalized by the Reynolds number at the orifice, for experiments D033, D034, D035, D036, D051 and D052.

The Probability Density Function (PDF) of the mass density of cases D033, D034, D035, and D036 at two locations along the jet's axis is shown in Figure 10 and Figure 11. One can see that the density at fixed $y/D$ exhibits one peak and a plateau towards smaller values. The peak represents the density of the environment, while the plateau the mixing region with the injected fluid. For all cases, the width of the PDF at $\frac{y}{D} = 80$ is smaller than at $\frac{y}{D} = 40$. This is an indication of better mixing further away from the jet's orifice. Also, the maximum peak at $\frac{y}{D} = 80$ is slightly smaller, indicating that the jet is better mixed with the environment. Cases D033 and D034 (Figure 10) are characterized by a positive density difference between the jet and the environment (positive buoyant jets), while cases D035 and D036 (Figure 11) by a negative one. We can observe that the width of the PDFs is larger for case D033 than for case D034 (Figure 10), and for cases D035 than for D036 (Figure 11). Indeed, cases D034 and D036 have a smaller Reynolds number than cases D033 and D035, and a larger magnitude of the Richardson number, indicating that a larger effect of buoyancy with respect to shear helps mixing.

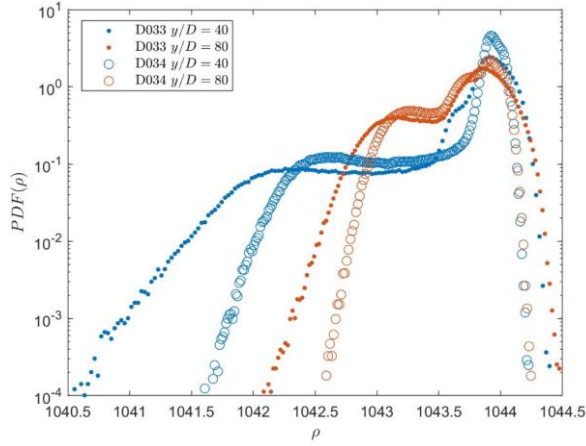 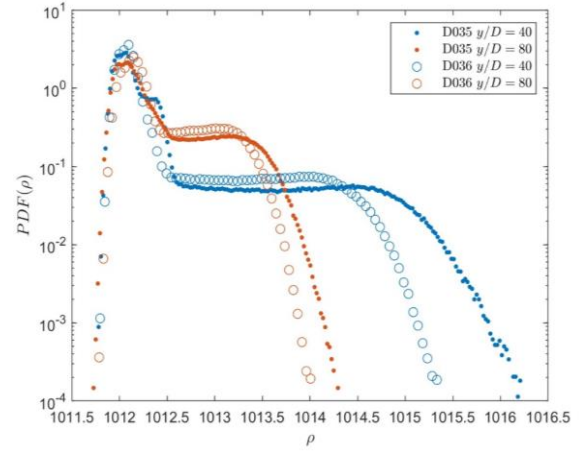

Figure 10: Probability Density Function (PDF) of the mass density of cases D033 and D034 at two locations along the jet's axis $\frac{y}{D} = 40$ (blue), and $\frac{y}{D} = 80$ (red).

Figure 11: Probability Density Function (PDF) of the mass density of case D034 at two locations along the jet's axis $\frac{y}{D} = 40$ (blue), and $\frac{y}{D} = 80$ (red).

The turbulent shear stress is defined as: $\tau(x,y) = \frac{\overline{u'(x,y,t) \cdot v'(x,y,t)}}{V^2(x_0,y)}$, where $\overline{u'(x,y,t) \cdot v'(x,y,t)}$ is the time average of the fluctuating velocity components $u'$ (horizontal) and $v'$ (vertical). The coordinates $(x,y,t)$ are the horizontal, vertical, and time coordinates shown in Figure 2, and $V(x_0,y)$ is the vertical velocity along the axis of the jet ($x = x_0$). Figure 12 shows the turbulent shear stress at the position of $\frac{y}{D} = 60$, for all the experiments and two references from the literature (Wygnanski [39] and Rodi [40]). We can see that all experimental cases and the two literature references well agree with each other, and the only exception is case D043, the one with stratified environment and lower Reynolds number.

Analogously to the turbulent shear stress, the turbulent concentration fluxes in the vertical and horizontal direction, $\tau_{vc}$ and $\tau_{uc}$ respectively, are defined as: $\tau_{vc} = \frac{\overline{v'(x,y,t) \cdot c'(x,y,t)}}{V(x_0,y) \cdot C(x_0,y)}$ and $\tau_{uc} = \frac{\overline{u'(x,y,t) \cdot c'(x,y,t)}}{U(x_0,y) \cdot C(x_0,y)}$, where $C(x_0,y)$ is the vertical concentration and $U(x_0,y)$ the horizontal velocity along the axis.

Figure 13 and Figure 14 show the turbulent concentration fluxes $\tau_{vc}$ and $\tau_{uc}$, respectively, at the axial position of $\frac{y}{D} = 40$, since the concentration was measured only in the bottom layer for the two cases with stratification. $\tau_{vc}$ and $\tau_{uc}$ are plotted as a function of the horizontal dimension $x/r_{1/2}$, where $r_{1/2}$ is the lateral dimension where the jet's vertical velocity is equal to the half of its maximum value. From Figure 13 and Figure 14, one can observe that no significant differences emerge among the experimental cases considered, except for the two jets with stratified environment (D042 and D043). For these two experiments we note that the turbulent concentration fluxes have a smaller magnitude in the central part of the jet and a wider distribution in the horizontal direction ($x$). This is an indication that the sharp stratification in the environment induces the jet to spread horizontally.

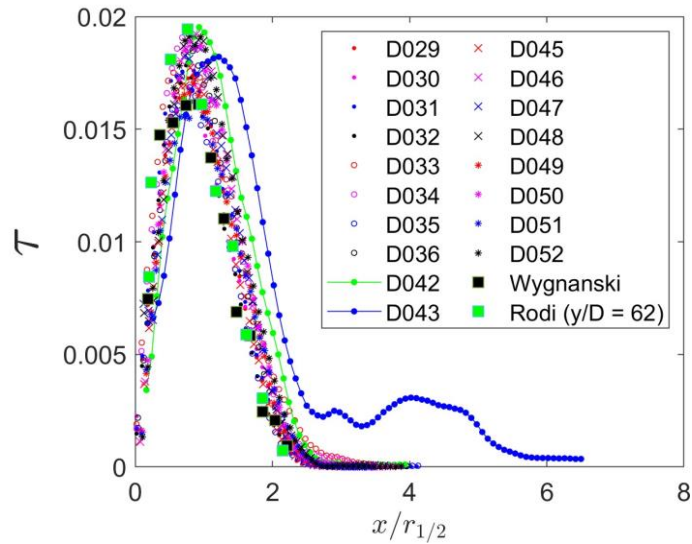

Figure 12: Turbulent shear stress at y/D = 60.

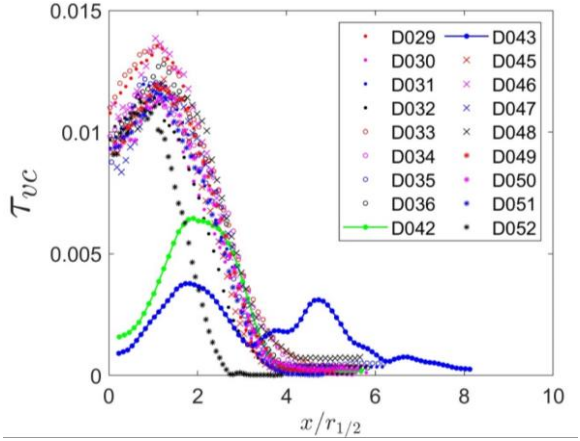 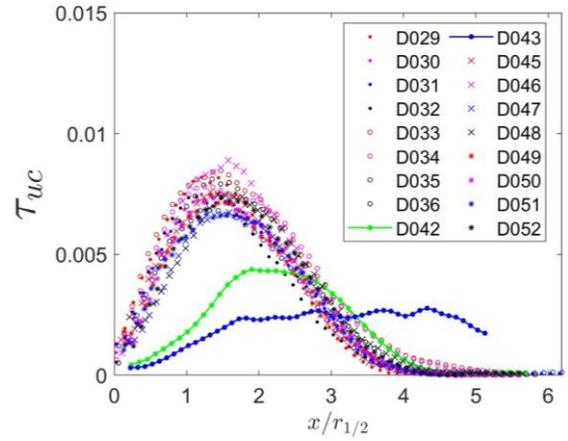

Figure 13: Vertical turbulent concentration flux at y/D = 40.

Figure 14: Horizontal turbulent concentration flux at y/D = 40.

Figure 15 shows the volume flux as a function of distance from the jet origin for all experimental cases. The volume flux is defined as the ratio between the entrainment ($\mu$) and the jet specific mass flux at the orifice ($Q = \rho_0 V_0 A$, where $\rho_0$ is the mass density and $V_0$ the jet's velocity at the orifice, while $A$ is the cross-section). The entrainment is defined as: $\mu(y) = \int_0^{L(x)} V(x,y) 2\pi x \, dx$, where $V$ is the time-averaged vertical velocity, and $L$ is the characteristic length of the jet. From Figure 15 one can observe that the volume flux is the lowest for the two jets injected in sharp environment, D042 and D043. For case D042, which has lower Reynolds number, the volume flux drastically decreases in correspondence with the position of the interface (about $\frac{y}{D} \simeq$ 60). One can also see that the cases with the largest Richardson numbers D034 and D052 have the lowest values of entrainment. The case with the largest volume flux is D048, which is characterized by a large negative Richardson number. From this comparison, one can conclude that a stratified environment and the Richardson number have a large influence on the volume flux and its evolution along the jet's axis. A stratified environment and a large positive Richardson number are responsible for lower values of volume flux. The same plot presents a reference slope from the experimental literature review of List [7] for comparison, which agrees well with the cases in uniform environment.

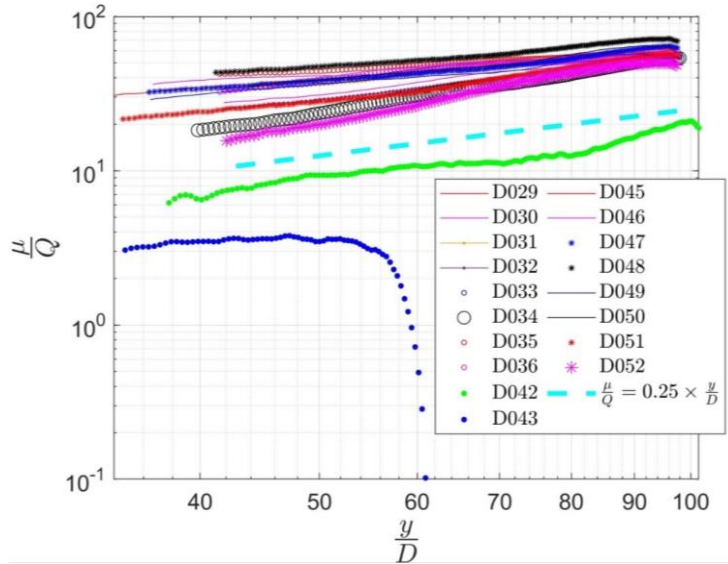

Figure 15: Volume flux as a function of distance from the jet origin for all experimental cases. Reference dashed line from [7].

**IV. CONCLUSIONS**

In this study, we analyze positive, negative, and neutral buoyant jets, both in a uniform and in a two-layer stratified environment. The investigation is based on simultaneous PIV and LIF measurements in the self-similar region of the jet. A novel RIM method is applied to perform accurate measurements with density differences up to 8.6%, while matching the refractive index of the mixed solutions. We observe that the two cases with a stratified environment show significant differences with respect to the other ones. The case in stratified environment and lower Reynolds number is characterized by a full width at half height that decreases after the sharp interface in the stratified environment, while the same quantity increases linearly for all other cases. For this experiment, also the maximum velocity decreases after the interface, while it increases for the other ones. The turbulent shear stress and the concentration fluxes show that the case in stratified environment with the lowest Reynolds number spreads horizontally at approximately the height of the interface, behaving like a fountain. We also observed that for the experiments of turbulent buoyant jets in uniform environment the value of the Reynolds number along the jet's axis shows a positive dependency on the inlet Richardson numbers. Also, the jet's entrainment is larger for negative Richardson numbers than for positive ones confirming that the ratio between the mass density of the jet and the environment has an influence on the entrainment of the jet.

This work can be extended in several directions. One regards the investigation of the effect of a linearly stratified environment on the buoyant jets. This is a closer representation than a two-layer environment to what occurs in real applications, like for example in the upper plenum of the hot pool of sodium-cooled fast reactors. A further direction consists in pushing the current limit of 8.6% density difference of the RIM method, by finding new possible combinations of aqueous solutions. The results of this study, in particular the turbulent fluxes of buoyant jests in uniform and stratified environment, will be used for the validation of commercial CFD codes.

**ACKOWLEDGEMENTS**

This work was supported by the Swiss Federal Office of Energy SFOE, which we thankfully acknowledge.